\documentclass[12pt]{article}

\usepackage{epsfig}

{\end{list}}
\newcounter{enumct}

\newcommand{\captive}[1]{\rule{5mm}{0mm}%
\begin{minipage}{150mm}\caption[small]{#1}\end{minipage}}
 
\begin{document}
\sloppy

\pagestyle{empty}

\begin{flushright}
DTP-98/84\\
October, 1998\\
\end{flushright} 

\begin{center}
{\LARGE\bf Glueballs : the Naked Truth}\\[8mm]
{\Large{M.R. Pennington}}\\[3mm]
{\it Centre for Particle Theory, University of Durham,}\\[1mm]
{\it Durham DH1 3LE, U.K.}\\[1mm]
{\it E-mail: m.r.pennington@durham.ac.uk}\\[7mm]

{\bf Abstract}\\[1mm]
\begin{minipage}[t]{140mm}
The current status of the accepted glueball candidates is reviewed.
The difference between {\it bare} and {\it dressed} hadrons is emphasised.
What two photon processes, both production and decay, have taught us about these
is discussed.
\end{minipage}\\[5mm]

\rule{160mm}{0.4mm}

\end{center}

\section{The Hadron Family}
Though the primary focus of this talk will be on glueballs, gluonic states
belong to the family of hadrons and one cannot talk about them without
recalling key features of the whole tribe.  Our picture of the hadron
world has not been static, but has evolved over time, in response to
both theoretical and experimental developments. Indeed, how our
view of glueballs has changed will be a key aspect of this talk.

Historically, the hadron world became simple 30 years ago, when it was
recognised that all the hadrons we knew were made out of quarks.
As we all learnt at our mother's knee, the world of  light hadrons
can be understood in terms of just three flavours of quark~: {\it up, down}
and {\it strange}. This naturally leads to mesons being grouped in nonets
for each allowed $J^{PC}$ quantum numbers. If we look at the nine
lightest vector mesons, for example, we see that these
are ideally mixed with the neutral members being
states made of either non-strange or strange quarks. We infer this from
the decays of the $\rho$, $\omega$ and $\phi$ mesons: the heavier $\phi$ 
decaying overwhelmingly into $K{\overline K}$. This picture works not only
for vector mesons, but tensors too. There the $f_2'(1520)$ is seen to decay
strongly to $K{\overline K}$ and less than 1\% of the time to $\pi\pi$,
despite the much greater phase-space for the latter.  This ideal picture
can be tested, as is well-known, by studying their photon interactions ---
but that is for later.

The simple quark model leads us to expect towers of such multiplets;
so dense that we can only realistically separate out individual
quantum numbers below 2 GeV in mass. All the multiplets we know of
are close to ideally mixed --- apart from the lightest pseudoscalar and scalar
mesons : the pseudoscalars because they are the Goldstone bosons
of chiral symmetry breaking and the scalars because of their role
as the Higgs sector of this same symmetry breaking.

Then 25 years ago along comes QCD. At first this simplifies matters
by explaining that mesons made of $q{\overline q}$ and baryons of $qqq$
are the simplest colour singlets, but it complicates life by requiring that glue
is an integral part of any hadron.  It is the gluonic cloud that turns
a current quark with a mass of just a few MeV into a constituent quark
of 300 MeV. So, while you and I are built of protons and neutrons
and so contain many trillions of quarks, most of our mass is made
of glue. Since glue is such an integral part of hadronic matter,
may be we don't need a quark to seed a gluonic cloud,
 but perhaps gluons can self-seed
and so produce constituent glue.
With this in mind, QCD predicts that there should be other
colour singlet states than $q{\overline q}$ and $qqq$ :
multiquark states of $qq{\overline {qq}}$, etc, hybrids of $q{\overline q}g$
and glueballs of $gg$ and $ggg$, etc.

Let us first consider the multiquark states.  Clearly they would belong to
bigger multiplets, 81's of light flavour. These were classified and their properties investigated
in bag models~\cite{mit}, with the prediction that there should be whole towers of these too
waiting to be uncovered. Indeed, there was the idea that, while meson interactions
naturally produced $q{\overline q}$ mesons, baryon-antibaryon processes
might readily form $qq{\overline {qq}}$ or baryonium states~\cite{baryo}. Indeed, twenty years ago,
a number of candidates were found, which later experiments showed to be mere
 statistical fluctuations.  Though mesons with unusually large couplings
 to baryon-antibaryon processes undoubtedly exist~\cite{oakden}, our view has changed.
 Potential model calculations
 by Weinstein and Isgur~\cite{wein} (and then by others~\cite{speth,barnes0}) showed that four quark systems only bind in very particular
 situations. Firstly, the system must include an $s{\overline s}$ pair
 and arrange itself into $K$--$\overline K$ or $K^*$--$\overline K^*$
 configuration, so that only isospin 0 and 1 states should occur.
 Moreover, binding only takes place if they form $0^{++}$ quantum numbers.
 This leads us to expect a very small class of multiquark states
 all associated with kaonic channels --- states with these characteristics exist.
 [As an aside, let me comment that why such non-relativistic notions should apply to light quark systems
 is not really clear to me, given that chiral dynamics is the primary determinant
 of low momentum pseudoscalar interactions, but perhaps that is just my problem.]
 
 Now let us turn to hybrids. You can think of such states as having $q$, $\overline q$ and $g$
 constituents. Or, as seems appropriate in Lund, you can think in terms
 of strings. Ordinary mesons are tied by a string ( a colour flux tube )
 joining a quark and antiquark together.  For excited hadrons, a stringy
 picture is clearly in order, but for the ground states the flux tube is fat
 and wide and less string-like.  Nevertheless, you can imagine that
 the string rotates adding angular momentum to the system. It is these
 excitations that correspond to hybrids~\cite{paton}.  Once again such states should come in towers,
 but towers for the most part with the same quantum numbers as ordinary 
 $q{\overline q}$ mesons. It is then, only by completing the $q{\overline q}$
 multiplets with given $J^{PC}$, that we can hope to identify any extra states.
  However, in certain situations they have uniquely
 identifiable quantum numbers.
  
Thanks to the development of high performance photon detection,
one can study peripheral processes like $\pi^- p\to (\pi^0\eta)n$, where
both the $\pi^0$ and $\eta$ are detected by their two photon decay.
Now, if resonances are seen in the $\pi^0\eta$ channel, then these must have
even spins if they are $q{\overline q}$ states. GAMS pioneered such studies~\cite{GAMS} and
found that in the 1400 MeV region there was a strong forward-backward
asymmetry, indicating important contributions with odd spin.  While GAMS
were not able to show that this enhancement had the phase variation
required of a resonance, the more recent 
BNL-E852 experiment has found just that~\cite{BNL}. Such a state,
with $1^{-+}$ quantum numbers, must be beyond the quark model.
From its decay pattern, 
this $\widehat\rho(1405)$ is more likely to be a $q{\overline q}g$ hybrid, than
a $qq{\overline qq}$ state.  Surely other hybrid states exist too~\cite{hybrids}.

Now let us turn to glueballs. At first, these are states made of constituent glue,
with no intrinsic quark content. Clearly, the world studied in experiments is one
with quarks. However (quenched) lattice theorists live in a world without quarks ---
indeed they find it difficult to switch quarks on. Consequently, lattice gauge
computations can teach us about a possible bare glueball spectrum.
It has long been known that these show (i) that a bound state spectrum exists,
(ii) that the lightest state is a scalar and (iii) the next is a tensor state
about 1.5 times heavier.
\begin{figure}[t]
\begin{center}
\mbox{~\epsfig{file=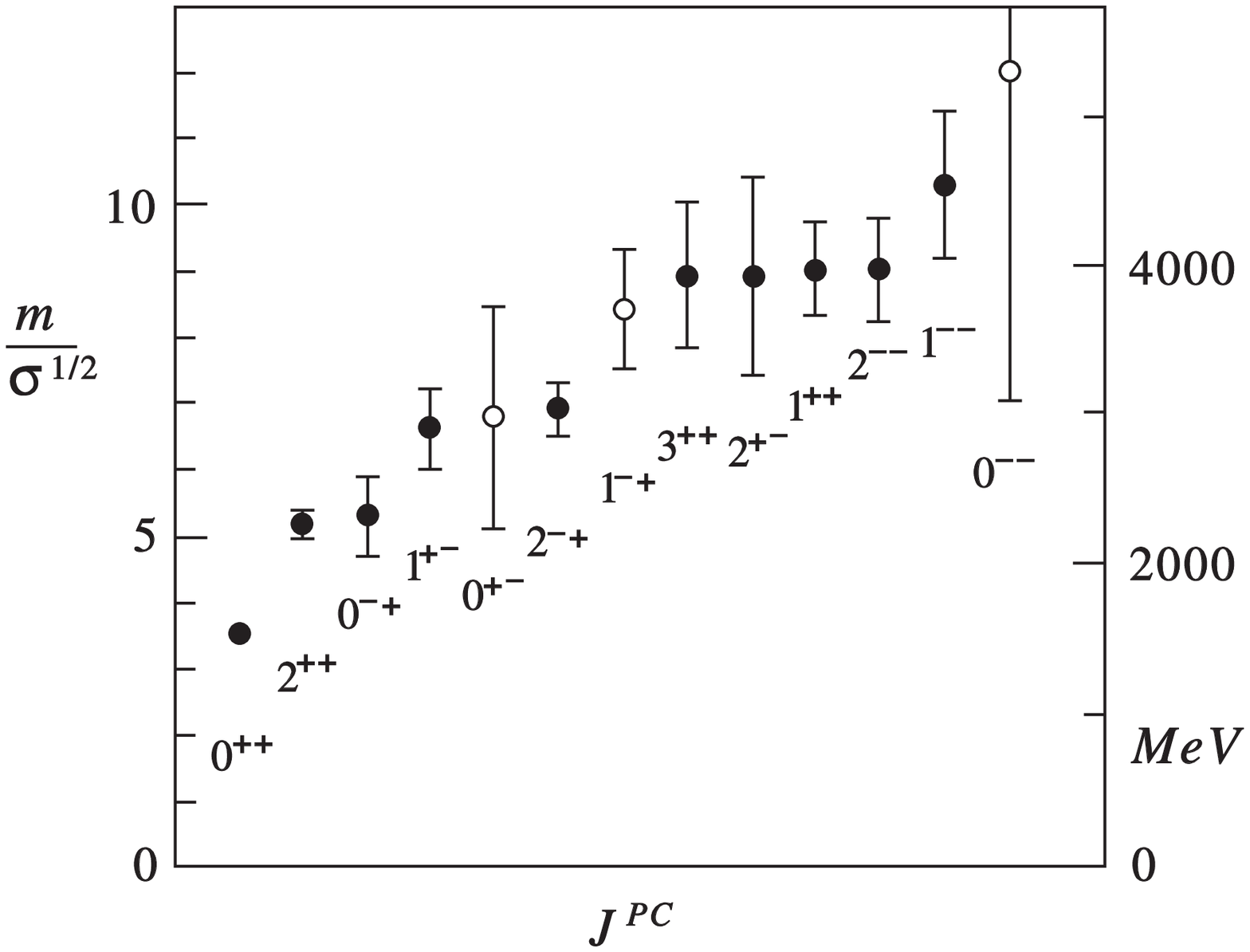,angle=0,width=11cm}}
\end{center}
\captive{Lattice results for the bare glueball spectrum in units of the
string tension $\sigma$ and in MeV (from UKQCD~\cite{UKQCD}).}
\end{figure}
In the last five years or so, these calculations have become much more precise,
see Fig.~1,
with these lowest two states having rather precise values in terms of the
{\it string tension}.  Estimating this for the pure glue world
from the spacing of the observed $q{\overline q}$
hadrons, masses for the lightest scalar state can be deduced.
UKQCD~\cite{UKQCD} quoted $(1568\pm 89)$ MeV, while the IBM group presented their GF11
result of $(1740 \pm 71)$ MeV~\cite{GF11}. Since, as we shall recall, there are
experimental candidates at masses of 1500 MeV and 1710 MeV, each claimed
support from these calculations, even though taking account
of their error bars these almost overlap. There then followed an interesting computation by Morningstar and
Peardon~\cite{mp}, using an {\it improved} action that gave $(1630 \pm 100)$ MeV.
The GF11 group have then re-evaluated their results~\cite{newibm} and 
 found the scalar mass to be $(1648 \pm 58)$ MeV~\cite{leewein}--- down from 1740 MeV.
 Indeed,  Weingarten calculated a {\it lattice} average of $(1632 \pm 49)$ MeV~\cite{leewein2}.
To the individual calculational groups, these differences  may
appear significant, but to the onlooker the remarkable achievement
is the consistency of these results summarised below!
$$
m_0(gg)\,=\,\cases {(1568 \pm ~89)&MeV\hskip 8mm UKQCD~\cite{UKQCD}\cr
(1740 \pm ~71)&MeV\hskip 8mm GF11~\cite{GF11}\cr
(1630 \pm 100)&MeV\hskip 8mm MP improved~\cite{mp}\cr
(1648 \pm ~58)&MeV\hskip 8mm GF11 reanalysed~\cite{newibm,leewein2}.\cr}
$$

\section{The experimental case for glueballs}
How do we learn about glueball states? The folklore is that one's best chance 
is in {\it so called} glue-rich processes. The best known example is $J/\psi$ decay.
There the $c\overline c$ quarks annihilate into gluons before creating 
lighter quark pairs to form the final state hadrons.  In $J/\psi$ radiative
decays, two gluons are the minimum number required by quantum numbers
and these form the basis for the lightest scalar and tensor glueballs, Fig. 2.
Other glue-rich reactions are double Pomeron processes. 
First let us consider
$p{\overline p}$ annihilation, particularly at rest.
There the idea is that the quarks in the initial proton and anti-proton annihilate
completely, leaving only glue, and this subsequently produces light hadrons.
The high efficiency of recent photon detectors has allowed all neutral final states to
be extensively studied at LEAR using the Crystal Barrel detector~\cite{amsler}. There
once more $\pi^0$'s and $\eta$'s
are identified by their two photon decay modes. Studying
$p{\overline p}\to (\pi^0\pi^0)\pi^0,\, (\pi^0\eta)\pi^0, (\eta\eta)\pi^0,\,(4\pi^0)\pi^0$ channels,
for example, the Crystal Barrel collaboration has  enormous statistics allowing not only the known
$f_2(1270)$, $a_2(1320)$ to be clearly picked out, but to study in great detail
the scalar states lying under these. These reveal~\cite{CB} the $f_0(1500)$ with a width of
$(112 \pm 10)$ MeV, as one of the best established states in the 
PDG tables~\cite{PDG}.
One should, however, be aware  that while some channels like $3\pi^0$
have been analysed allowing for both $S$ and $P$--wave annihilation at rest,
others like $5\pi^0$ have been treated with the restrictive assumption
that only $^1S_0$ is all that matters. Nevertheless, the ability of Crystal
Barrel to show consistency among a wide range of final state channels is impressive
and the only way to do such spectroscopy.  The $f_0(1500)$ has been clearly
established by Crystal Barrel and confirmed in charged pion channels by
 Obelix~\cite{ob}.
Moreover, it may well be the same state
as the $G(1590)$ found by GAMS~\cite{g1590} in the  $\eta\eta$ channel and the same as the
scalar signal found by LASS under the $f_2'(1525)$ in their 
${\overline K}K\to {\overline K}K$ data~\cite{lass}

Historically, $J/\psi$ radiative decay was the first to throw up serious 
glueball candidates.  Initially Crystal Ball at 
SPEAR~\cite{cbslac}, and
later Mark III~\cite{mkiii}, not only saw the $f_2'(1525)$ in both
the $K^+K^-$ and $K_S K_S$ decay modes, but a large signal at 1710 MeV, known as the $\theta$.
Preliminary analysis revealed this to be a tensor state~\cite{cbslac} and this was
claimed to be a glueball --- the lowest $q{\overline q}$ nonet being full
and 1710 MeV being too light for the radial recurrences. Subsequent analysis
suggested that the $\theta$ signal was largely $K{\overline K}$ $S$--wave~\cite{j0}.
A more measured  appraisal led to the proposal that it was a mixture of $J=0$ and 2,
and the state is now known as the $f_J(1710)$~\cite{PDG}. The same Mark III experiment
also found a narrow spike, dubbed the
$\xi(2230)$, in  $K{\overline K}$ channels~\cite{ximkiii}. 
However, this was not confirmed by the DM2 data, that 
clearly showed the $\theta$~\cite{dm2}. In the last couple of years the BEPC machine in China
as yielded interesting results on $J/\psi$ decays. The BES detector has found~\cite{xibes}
narrow spikes in the 2230 region, not only in the kaon channels,
but in $\pi\pi$ and $p{\overline p}\;$ too. Though each individually has
limited statistics, the fact that a narrow enhancement is seen in
all these channels, at the same mass, increases their significance.
If the $\theta$ is a scalar, then the tensor glueball slot is {\it freed}. The
$\xi(2230)$ is claimed to fill this space~\cite{gluebes}, having  a mass in agreement with lattice
expectations, Fig.~1.
  
\begin{figure}[b]
\begin{center}
\mbox{~\epsfig{file=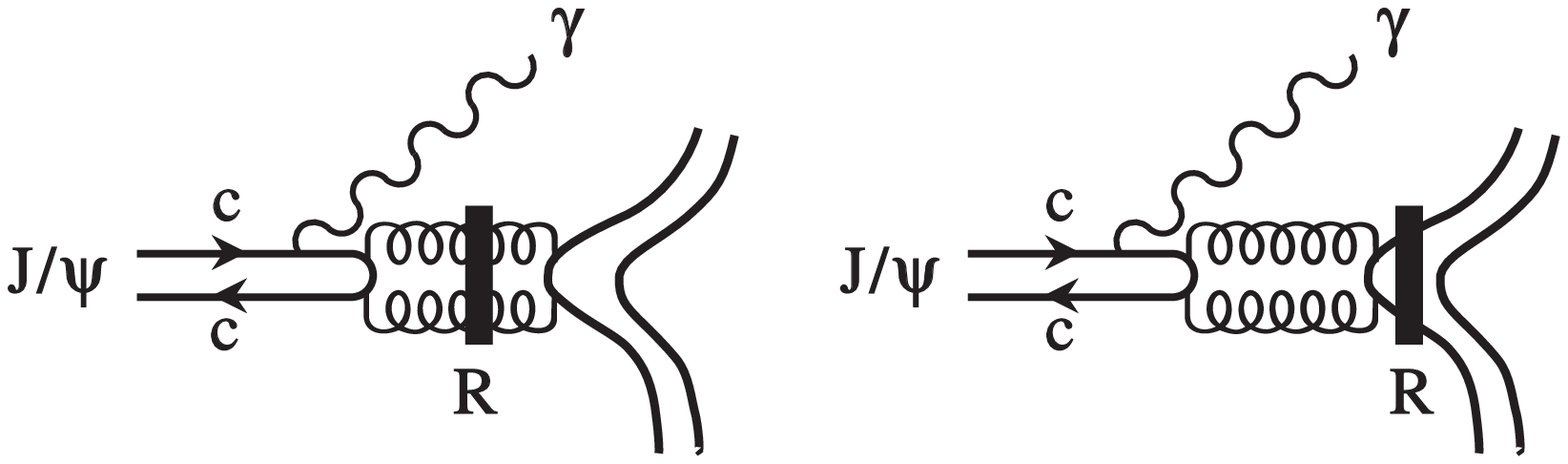,angle=0,width=13cm}}
\end{center}
\captive{Resonance formation in $J/\psi$ radiative decays in QCD perturbation
theory.}
\end{figure}

It is important to note that in $J/\psi$ radiative decays, ordinary $q{\overline q}$
states are also seen, like the $f_2(1270)$ in the $\pi\pi$ channel and
the $f_2'(1525)$ in the $K{\overline K}$. So how do we tell whether a state
is rich in glue, beyond it being extra to an established quark model multiplet?
An interesting test has been proposed by \c Cakir and Farrar~\cite{cf},
which has been applied to current phenomenology
in collaboration with Close and Li~\cite{closef}. As seen in Fig.~2, perturbatively 
we may think
of a resonance $R$ being either made of glue that binds and then subsequently
decays by creating $q{\overline q}$ pairs, or the glue may first create a $q{\overline q}$ pair,
which resonates, and then subsequently decays into ordinary hadrons.
Now if the state $R$ is glue-rich, it should be easy to produce, while if it
is a quark state then the gluons have to produce a $q{\overline q}$ pair first.
This latter process is perturbatively suppressed by $\alpha_s^2$ in rate
--- suppressed if the relevant momentum scale is set by the charmed quark mass.
(Of course, the $q{\overline q}$ pairs that have to be created to allow decays
into light hadrons
are assumed, as usual, to cost nothing, since a {\it strong} strong interaction
is needed.) Thus we would expect BR$(R\to gg)$ to be $\sim 100\%$ for a glueball
and only $\sim 10\%$ for a quark state. By comparing the rate for $J/\psi\to\gamma R$
to the inclusive decay rate, this gluonic branching ratio can be determined ---
if we buy the perturbative underpinnings.
Using the latest PDG98 values~\cite{PDG} for $J/\psi$ decays, one can deduce,
 following Close {\it et al.}~\cite{closef},
\begin{eqnarray}
\nonumber
{\rm BR}(f_2(1270)\to gg)&\simeq& (22 \pm 4)\%\ ,\\ \nonumber
{\rm BR}(f_2'(1525)\to gg)&\simeq& (17 \pm 4)\%\ ,
\end{eqnarray}
just as expected.

Now let us apply the \c Cakir-Farrar test to the glueball candidates we've discussed:
first, the $f_J(1710)$. The PDG98 tables~\cite{PDG} quote
$${\rm BR}(J/\psi\to \gamma f_J\to \gamma K{\overline K})\,=\,
\left(8.5^{+1.2}_{-0.6}\right)\, 10^{-4}\; .\eqno (1)$$
With the $f_J$ branching ratio to $K{\overline K}$ being between 50 and $100\%$,
one finds \cite{closef} 
\begin{eqnarray}
\nonumber
0.15 &<&{\rm BR}(f_2(1710)\to gg)\,  <\, 0.3 \quad{\rm if}\; J=2\; ,\\ \nonumber
0.52 &<&{\rm BR}(f_0(1710)\to gg)\,  <\, 1.0 \quad{\rm if}\; J=0\; .
\end{eqnarray}
Close {\it et al.} emphasise that this means that a scalar 1710 MeV state must be
glue-rich.  However, BES have something rather different to say.
They are the first group to present a separate
branching ratio for the $J=0, 2$ parts of  their $f_J$ signal.
They have an $f_2(1696)$ and an $f_0(1780)$, each with a width of about 100 MeV~\cite{fjbes}.
They then quote
\begin{eqnarray}
\nonumber
{\rm BR}(J/\psi\to \gamma f_2\to \gamma K{\overline K})&=&
(2.5 \pm 0.4)\, 10^{-4}\;\\ \nonumber 
{\rm BR}(J/\psi\to \gamma f_0\to \gamma K{\overline K})&=&
(0.8 \pm 0.1)\, 10^{-4}\;.
\end{eqnarray}
Since for the scalar, this branching ratio is only a tenth of the PDG98 average,
Eq.~(1),
it would make their scalar have less than a 10\% branching ratio to $gg$.
Why the BES rates are so different from the results~\cite{PDG} summarised
in Eq.~(1) is a pressing problem.

A similar inconsistency unfortunately applies to the $f_0(1500)$ and
the \c Cakir--Farrar test. Firstly, the $f_0(1500)$ had not
been identified in the experimental results on $J/\psi$ radiative 
decays. Indeed, Mark III originally found a sizeable $0^-$ wave in
the 1.5 GeV mass region.
Nevertheless, from the Crystal Barrel results we know the $f_0(1500)$ exists with a large
$4\pi$ decay mode.  This  prompted Bugg and collaborators together with
Burnett~\cite{psibugg} to go back and re-analyse the $J/\psi\to \gamma (4\pi)$
data from Mark III, building in the $f_0(1500)$ state. From this analysis
they find
$${\rm BR}(J/\psi\to \gamma f_0(1500)\to \gamma 4\pi)\,=\,
(5.7 \pm 0.8)\, 10^{-4}\;.$$
Assuming the $4\pi$ decay mode of the $f_0(1500)$ is less than 50$\%$,
in keeping with the Crystal Barrel results, Close {\it et al.} find
$${\rm BR}(f_0(1500)\to gg)\, > \, 0.9 \pm 0.2\; ,$$
once again signalling a large gluonic component. However,
BES give results for the $\pi^0\pi^0$ channel,
 from the fits to which one might infer (see Fig. 2 of ~\cite{bespipi}):
$${\rm BR}(J/\psi\to \gamma f_0(1500)\to \gamma \pi^0\pi^0)\,\simeq\,
(4 \pm 2)\, 10^{-5}\;.$$
With the $\pi^0\pi^0$ rate from the LEAR experiments~\cite{PDG}, one gets
$${\rm BR}(f_0(1500)\to gg)\, \simeq \, 0.4 \pm 0.2\; $$
instead. Disappointingly, a factor of 2 in disagreement.
To clear this up we need a good statistical sample of
$J/\psi\to \gamma(2\pi)$ from BES 
or $\Upsilon\to \gamma(2\pi)$ from CLEO or a B--factory to analyse.
If this is not possible, then a large $4\pi$ data-set might suffice.
Particularly useful would be the $4\pi^0$ final state, because of its freedom
from a $\rho\rho$ contribution that dominates charged four pion channels
in this mass range (as Close {\it et al.}~\cite{closef} have stressed).

\begin{figure}[h]
\begin{center}
\mbox{~\epsfig{file=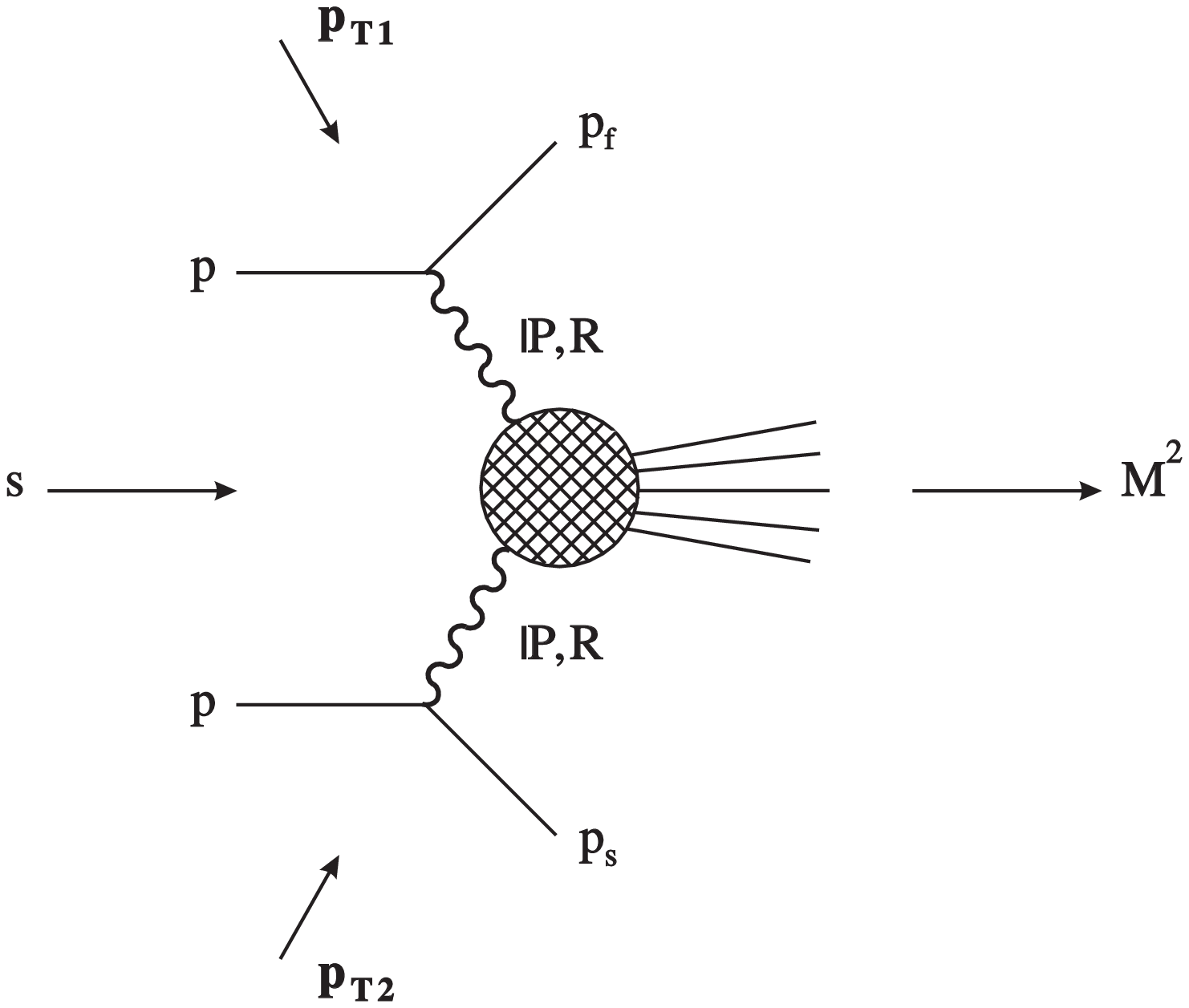,angle=0,width=10cm}}
\end{center}
\captive{Central production of hadrons in $pp\to ppX$, by Pomeron and
other Regge exchanges.}
\end{figure}

Let us now turn to central production, Fig.~3. Recall that the dominant soft
 scattering mechanism
at high energies in $pp$ collisions can be pictured as : 
each proton is a collection of quarks surrounded by a cloud of gluons.
As they approach each other in the c.m. frame, a colour singlet bit of glue gets detached from one
proton and is absorbed by the other. This we call Pomeron exchange.
Now once in a while, as the two protons approach, colour singlet glue
from each will be released and these bits may fuse to produce resonance states.
These would be preferentially gluonic in nature. The state would decay into
hadrons observed in the central region of rapidity, well separated from the on-going protons.
Of course, not only the Pomeron with its vacuum quantum numbers need be 
exchanged, but other Reggeons can too, producing conventional hadrons. It has been understood for thirty years
that the Pomeron contribution dominates at very small momentum transfers
between the protons.  This was the idea underlying the ISR experiment with the Axial Field
Spectrometer and its {\it Roman pots}~\cite{afs}.  However, there only limited final states
could be studied, like $\pi^+\pi^-$.  More recently, a whole series of experiments
have been performed at CERN with the Omega spectrometer.
With results~\cite{wakirk} from WA76, 91 and 102 on central production at 
85, 300 and 450 GeV/c (in the lab. frame) with a wide range of final states, 
the systematics of these  processes has become
possible. An intriguing systematic relation has been noted by
 Close and Kirk~\cite{closek}.

\begin{figure}[th]
\begin{center}
\mbox{~\epsfig{file=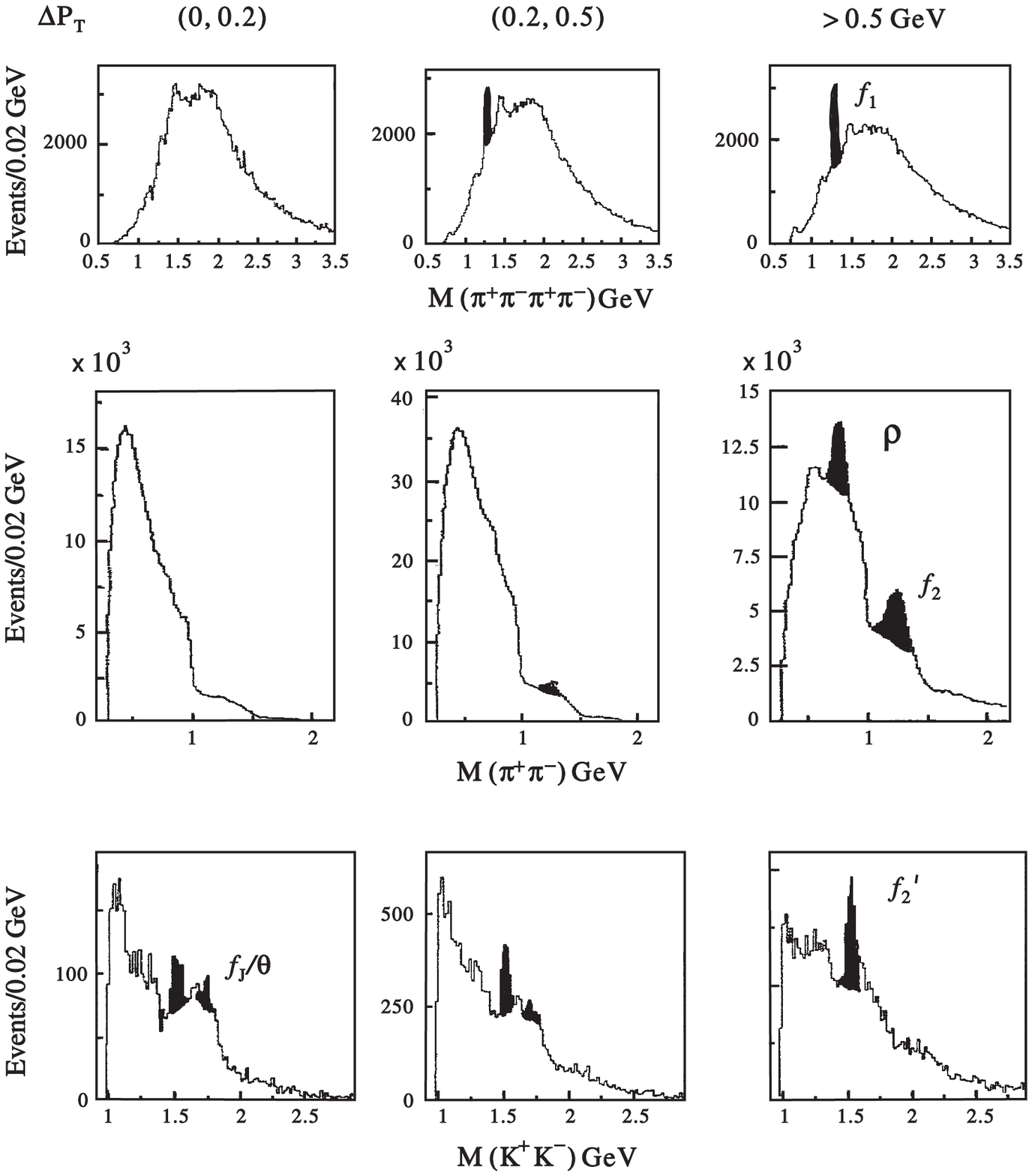,angle=0,width=13.5cm}}
\end{center}
\vspace{-3mm}
\captive{$\Delta p_T$ dependence of central production of three typical
hadronic final states in $pp\to ppX$ from WA102 \cite{wakirk,closek}.}
\end{figure}

First one defines
the relative tranverse momentum between each pair of initial and final protons,
i.e. $\Delta p_T\,=\,\mid p_{T1} - p_{T2} \mid$, Fig.~3. The data are binned, for each
exclusive final state, into $\Delta p_T$  (a) $ < 200$ MeV/c, (b) 
$\in (200,500)$ MeV/c, and (c) $ > 500$ MeV/c. As seen in Fig.~4,
conventional $q{\overline q}$ states dominate at {\it larger} $\Delta p_T$.
In the $\pi^+\pi^-\pi^+\pi^-$ channel, the $f_1(1285)$ is beautifully seen
with $\Delta p_T > 500$ MeV/c, less so at intermediate values and then
essentially disappearing at {\it small} $\Delta p_T$. A similar effect is observed
with the $\rho$ and $f_2$ in the $\pi^+\pi^-$ channel and with the $f_2'$ in
$K^+K^-$. In contrast, the $\theta/f_J$ appears more prominently
at small $\Delta p_T$.  Why this occurs is matter of conjecture.
Nevertheless, Close and Kirk~\cite{closek} have shown that states, that are not
obviously members of simple $q{\overline q}$ multiplets, are produced
at smaller values of relative transverse momentum.  This is a valuable way of 
enhancing the signal to background for such {\it unusual} states.
Some day we may understand why this works.

\section{Bare states and dressed hadrons}
Now let us look at the spectrum of scalars and tensors that we have been discussing
with an eye to their glueball candidacy. 
 In Fig.~5 are shown the states in the PDG98 tables~\cite{PDG}.
 
\begin{figure}[h]
\begin{center}
\mbox{~\epsfig{file=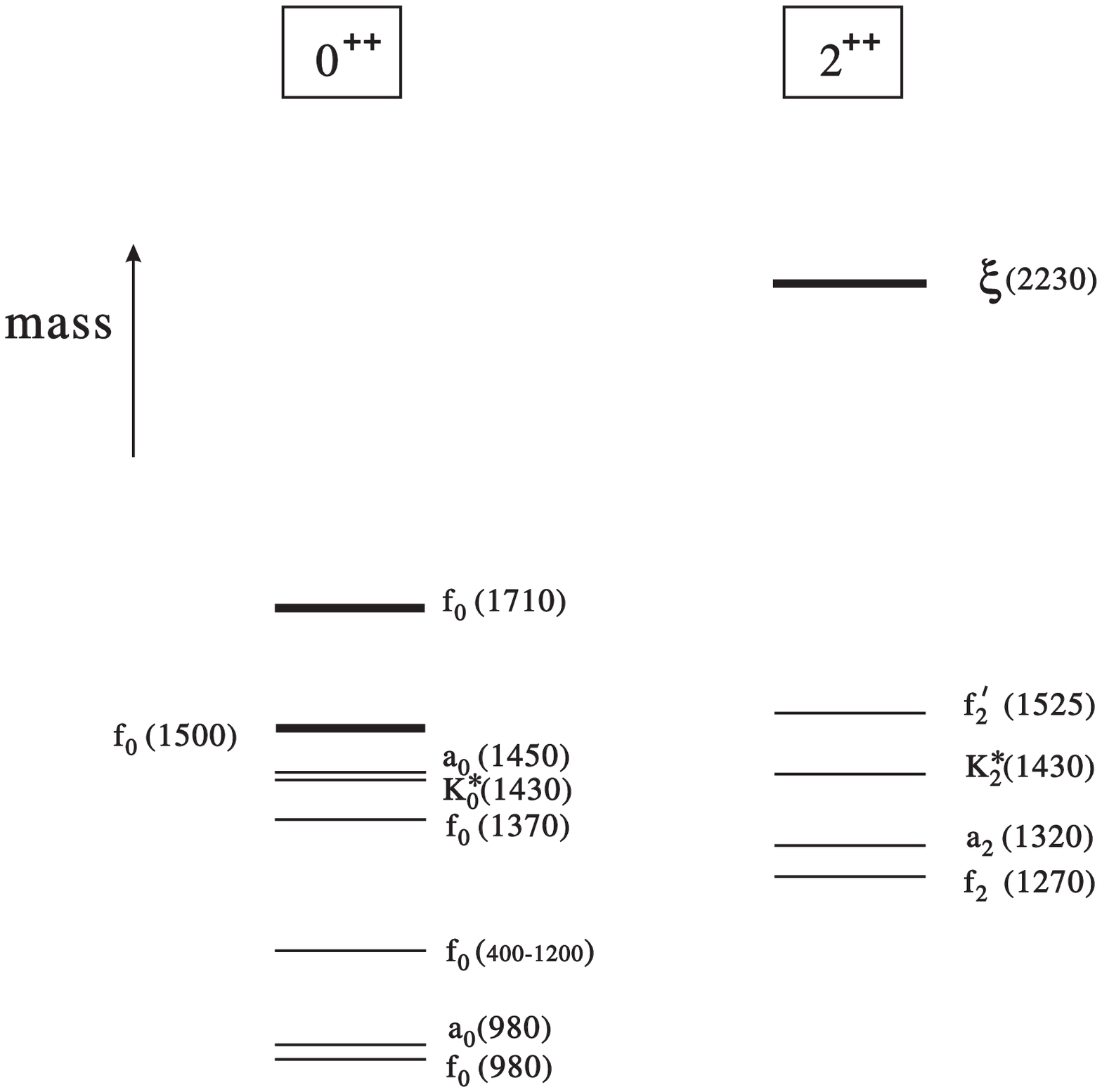,angle=0,width=13.cm}}
\end{center}
\captive{Mass spectrum of the lightest scalars and tensors from the
 1998 PDG tables. The bold lines indicate the glueball candidates.}
\end{figure}

The lattice suggests that the bare scalar glueball is around  1600 MeV,
and the tensor is near 2300 MeV. With this in mind,
what is immediately apparent  from Fig.~5 is that the tensor $q{\overline q}$ nonet is well separated
from the $\xi(2230)$, a possible  glueball. In contrast, right among the
many scalar states are the potentially gluish $f_0(1500)$ and $f_0(1710)$.
This inevitably means that the scalar glueball will mix with the nearby 
$q{\overline q}$ states. This gives a simple explanation of why the scalar
glueball candidates have widths of 100 MeV or so, while the $\xi(2230)$
 may be only 20 MeV wide~\cite{xibes}.
Clearly, for the scalars we have to take mixing into account.
Let me discuss two of the ways in which this has been done.

\begin{figure}[b]
\begin{center}
\mbox{~\epsfig{file=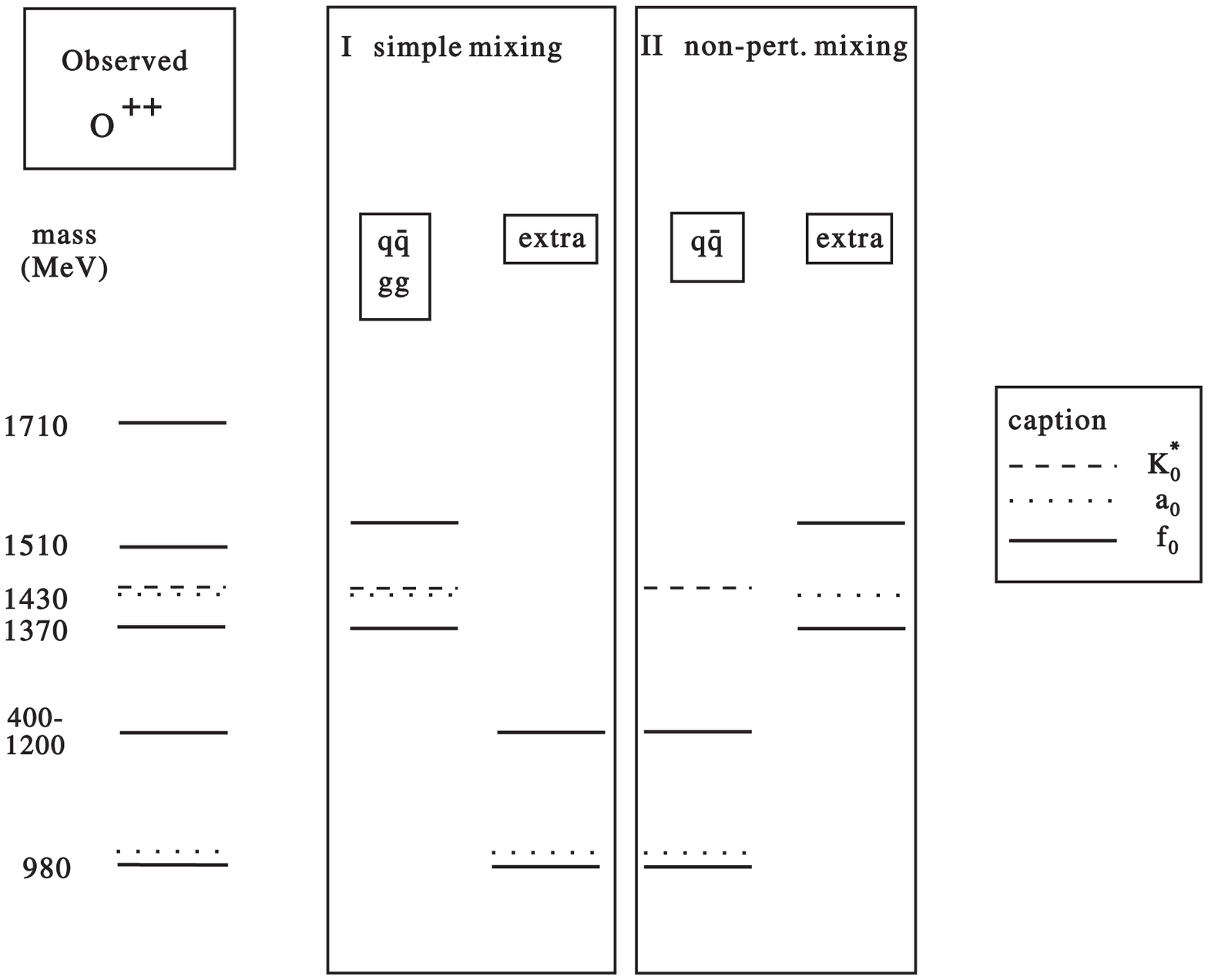,angle=0,width=14.2cm}}
\end{center}
\captive{The spectrum of light scalars and how they are described in two
possible mixing schemes : I from non-relativistic first order
perturbation theory by Amsler and Close \cite{amslerc}, II from non-perturbative hadron dressing
by Tornqvist \cite{nils}.}
\end{figure}

The first based on non-relativistic perturbation theory
is the simplest to understand. Amsler and Close~\cite{amslerc}
pick out the ten scalars shown in the {\it simple mixing} column in Fig.~6 and presume they
are formed from a quark model nonet and a glueball. 
As usual, we denote the isoscalar  $u{\overline u}$, $d{\overline d}$ combination
 by $n{\overline n}$ and that with hidden strangeness
by $s{\overline s}$.
If the unmixed glueball is between these two members of the nonet,
 then first order perturbation theory makes the $f_0(1500)$ largely gluonic and the
 $f_0(1710)$ largely one with hidden strangeness and consequently have a large
 $K{\overline K}$ decay mode. In contrast, Lee and Weingarten~\cite{leewein,leewein2} place the
 unmixed glueball above both the $n{\overline n}$ and $s{\overline s}$ states. They then find
 the $f_0(1710)$ to be largely gluonic and the $f_0(1500)$ mainly $s{\overline s}$.
 [Unfortunately, it is quite unclear, at least to me, why non-relativistic first order perturbation
  theory should apply to a bound state system of light quarks
  where by definition the interactions must be relativistic and strong.]
   As indicated in Fig.~6,
  this scheme leaves the $f_0$, $a_0(980)$ out in the cold, as extra states
  to be explained some other way~\cite{mit,wein,barnes0,achasov}.
  
  A strong physics approach to this problem has been proposed by Tornqvist~\cite{nils}.
  The basic idea is that the quark model applies to non-interacting bound 
  states --- states that don't decay: bare states. Thus the bare $\rho$ is a 
  bound state of $u$
  and $d$ quarks. The physical $\rho$ decays to $\pi\pi$. It does this by spending some
  of its time in this multiquark configuration. Such configurations dress 
  simple bound states turning them into hadrons.  In the case of the vectors and tensors, there
  is a very close connection between the underlying bare states of the quark model
  and the hadrons we observe.  However, as has been emphasised by 
  Tornqvist~\cite{nils} and by 
  Bugg~\cite{buggcors}, in particular, this is not the case for scalars. The dressed and bare states
  are quite different. 

To see why, we need to invoke a non-perturbative formalism for discussing this problem.
  As an illustration consider the lightest vector  meson multiplet.
  The nine underlying $q{\overline q}$ states are ideally mixed with a mass splitting induced by the
  strange quark being 100-120 MeV heavier than the $u$ and $d$ quarks.
  Such bare states with a mass $m_0$ have propagators with denominators that
  are just $m_0^2 - s$. They have poles on the real axis in the complex $s$--plane.
  Such states have no decays --- that's the first term on 
  the right hand side of Fig.~7.
   One then turns on the interactions, in which
  the $\phi$ can couple to $K{\overline K}$ and the $\rho$ to $\pi\pi$, for instance.
  The effect of these interactions can be summed,
   as shown long ago by Dyson, to give the propagator
  illustrated in Fig.~7. 
\begin{figure}[h]
\begin{center}
\mbox{~\epsfig{file=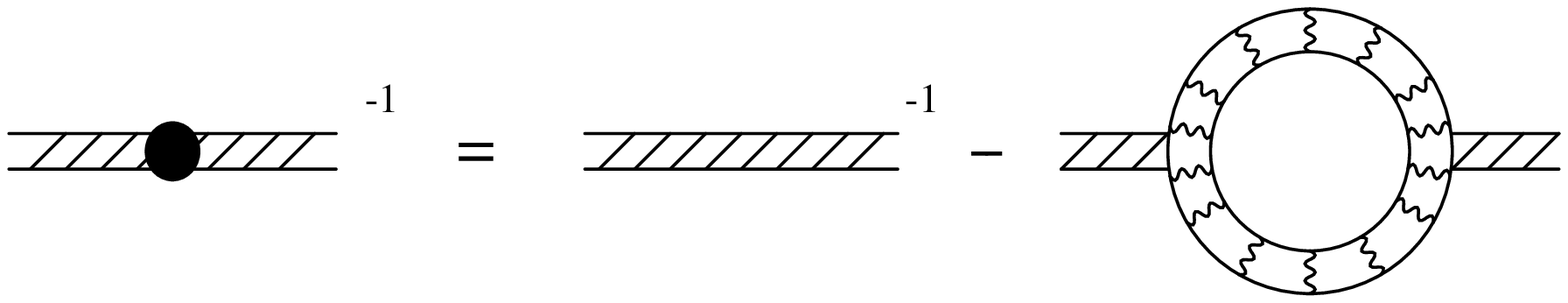,angle=0,width=11cm}}
\end{center}
\captive{The bare bound state propagator is dressed by hadronic
interactions. The dot signifies the dressed propagator. The loop is of 
$q \overline q$ mesons, predominantly pseudoscalars: the wiggly lines are to emphasise these too
are bound states.}
\end{figure}

\noindent This gives the inverse  propagator of the dressed hadron.
  This has a form that is no longer simply $m_0^2-s$, but is
  $\;{\cal M}(s)^2-s-i{\cal M}(s)\,\Gamma(s)$. The pole is now in the complex
  plane, below the cut produced by each particle threshold. This pole now
   corresponds to a decaying particle.
  How far the pole moves depends on the
  strength of the interaction with the hadrons to which it couples, Fig.~7.
  The dominant intermediate states (the particles in the loops) are two
  pseudoscalars --- this applies to scalars as well as vectors and tensors.
  For vectors, the strength of the interaction is fixed by the
  coupling of the $\rho$, for example, but being a $P$--wave interaction
  the discontinuity across the cut is suppressed by the usual threshold factors.
  Consequently, the resulting hadron is very close to the underlying quark model state.
  For instance, while the bare $\phi$ is purely $s{\overline s}$, the dressed hadron
  has a Fock space that is
$$\vert \phi \rangle\;=\;\sqrt{1-\epsilon^2}\,\vert s{\overline s}\rangle\;+\;
\epsilon_1 \vert K{\overline K} \rangle\,+\,\epsilon_2 \vert \rho\pi \rangle\,
+ ...\; ,$$
where $\epsilon^2 = \epsilon_1^2 + \epsilon_2^2 + ...$.
 With $\epsilon^2 \ll 1$, the physical $\phi$ is overwhelmingly an $s{\overline s}$
state, just like the bare one, and the switching on of interactions 
has produced a relatively small effect.  Consequently,
 the simple quark picture works.
 
 However, the case for scalars is quite different. Whilst the same dressing
 pictorially applies, Fig.~7, its magnitude is far greater. The scalars
 couple 2 or 3 times more strongly to pseudoscalars. Moreover, their
  interactions are $S$--wave, making the effect of the thresholds much more
  pronounced~\cite{buggcors}. Tornqvist has turned these words into a calculational scheme~\cite{nils}.
  The underlying quark multiplet is assumed to be ideally mixed.
  Its mass is fixed by the requirement that the dressed strange states
  become the $K_0^*(1430)$~\cite{PDG}. This fixes the position of the
  underlying bare multiplet. The undressed isotriplet $n{\overline n}$
  state is then at 1420 MeV. However, so strong is the hadronic dressing, that the
  state moves down to $K{\overline K}$ threshold and becomes the $a_0(980)$.
  Indeed, while the bare $\vert a_0(1420) \rangle_0\,=\,\vert n{\overline n} \rangle$,
  the dressed hadron this scheme predicts is:
  $$\vert a_0(980) \rangle_1\,=\, (0.2)^{1/2} \vert n{\overline n} \rangle\,
  +\, (0.7)^{1/2} \vert K{\overline K} \rangle\,+\,
  (0.1)^{1/2} \vert \pi\eta' \rangle, \eqno(2)$$  
  remarkably like the $a_0(980)$ of the PDG~\cite{PDG}.
  (The subscript \lq 1\rq~ 
  indicating that only two pseudoscalar interactions
  have been included in calculating the dressing in Fig.~7.)
  
  In a similar way, the isoscalars at 1420 MeV and its $s{\overline s}$ partner at 1620 MeV
  produce a broad $f_0(1200)$ and a narrower $f_0(980)$.
  The latter, like the $a_0(980)$, has a large $K{\overline K}$ component, just as experiment requires.
  Thus, in the scalar sector the underlying quark model states and the hadrons we observe
  are not  simply related. This scheme accounts for the nine lightest observed
  scalars as resulting from an underlying quark model nonet. It leaves the
  $f_0(1500)$, and  possible $f_0(1370)$, $a_0(1450)$ and $f_0(1710)$ 
  to be explained (Fig.~6).
  
  Elena Boglione and I extended this Schwinger-Dyson approach to
  include a bare glueball~\cite{elenaglu}. As we have seen, the lattice delivers this
  at 1640 MeV or so. Indeed, the GF11 group have calculated  its
  coupling to two pseudoscalars, finding that naively this would correspond to a width 
  of $(108 \pm 29)$ MeV~\cite{GF11}. Moreover, they find that the couplings to
  $\pi\pi$, $K{\overline K}$ and $\eta\eta$ are not those expected of an $SU(3)_f$ singlet, but rather
  the heavier pseudoscalars are favoured~\cite{GF11,leewein}. Such a state will mix through its
  common decay channels with the $q{\overline q}$ scalars. Boglione and I found that this had little
  effect on the dressed $q{\overline q}$ scalars just discussed. However, while the
  glue state is moved downwards 20-30 MeV or so in mass, its width to 
  pseudoscalars decreases dramatically --- $\eta\eta$ channel is still favoured, but the width to
  this is only $\sim 20$ MeV. Though this is not unlike the $f_0(1500)$
  of Crystal Barrel, its total width is too small. This shows the importance
  of including multipion channels in such a calculation if one is to get
  out reliable (and realistic) predictions for states in the 1.5 GeV region.
  Steps in this direction have been made~\cite{elenaglu,anisovich}.
  
 \section{Glueballs as seen by photons}
Resonance production in two photon reactions, as in $e^+e^-\to e^+e^-R$ (Fig.~8),
is, in principle, a very clean way of learning about the structure of the
hadron $R$.
\begin{figure}[th]
\begin{center}
\mbox{~\epsfig{file=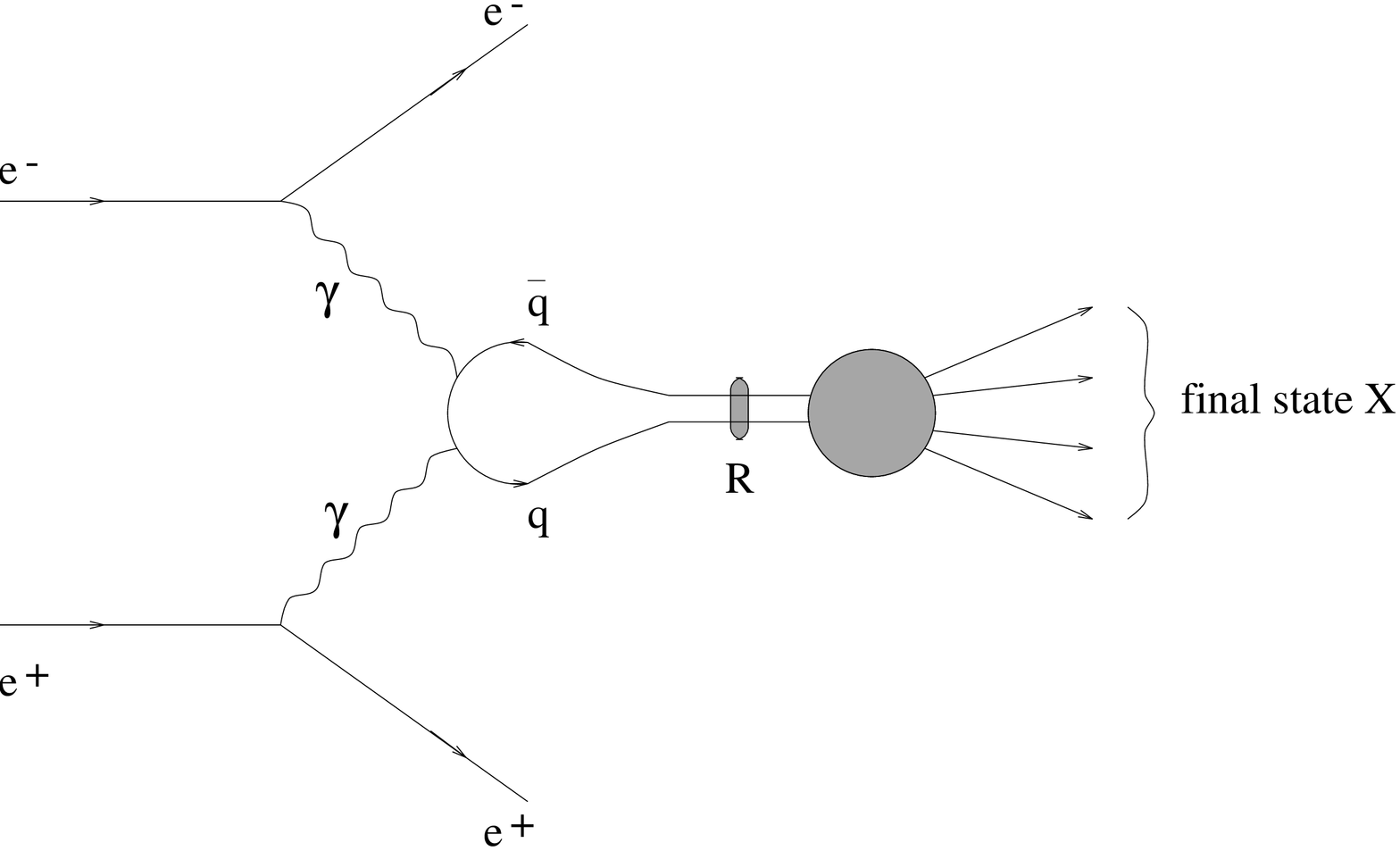,angle=-90,width=11.4cm}}
\end{center}
\vspace{-11mm}
\captive{Two photon production of a $q{\overline q}$ resonance in 
$e^+e^-\to e^+e^-X$.}
\end{figure}
 The photons couple to the electric charge of the constituents. 
 Tensor states readily couple to two photons, so these are the easiest to consider.
 As seen from Fig.~8, the rate to two photons is given by

 $$\Gamma(2^{++}\to\gamma\gamma)\,=\,\alpha^2 \langle\
  e_q^2\ \rangle^2\;\Pi_R\quad, \eqno (3)$$
 involving the square of the average constituent charge squared times
 the probability that the constituents annihilate, $\Pi_R$.
 Assuming that this probability is similar for members of the same quark multiplet,
 one obtains the well-known relation that
 $$\Gamma(f_2\to\gamma\gamma)\,:\, \Gamma(a_2\to\gamma\gamma)\,:\,
 \Gamma(f_2'\to\gamma\gamma)\,=\, 25\,:\, 9\,:\, 2\quad ,$$
 for an ideally mixed multiplet.  Experiment~\cite{PDG} gives $25 : 10 \pm 2 : 1\pm 0.2$.
 The $f_2$ radiative width is what sets the scale, so I should discuss how well
 we really know this. 
 
 But before this
 we should note that a naked glueball would not be seen in two photon reactions.
This is the motivation for {\it stickiness} that Chanowitz first introduced~\cite{chanowitz}.
Since a glueball should appear strongly in $J/\psi$ radiative decay (Fig.~2), but very weakly in
$\gamma\gamma$ reactions (Fig.~8), he defines the quantity {\it stickiness} ${\cal S}$
for a generic hadron $h$ by
$${\cal S}\,=\, C\, \left({M(h)}\over{k_{\gamma}}\right)^{2\ell+1}\;
{\Gamma(\psi\to \gamma h)\over{\Gamma(h\to\gamma\gamma)}}\quad ,$$
where $\ell$ is the lowest orbital angular momentum needed for state $h$ to
couple to two vectors. Simple glueballs should then have a large element of stickiness. With the normalisation $C$ chosen so that
${\cal S}=1$ for the $f_2(1270)$, CLEOII~\cite{CLEO} find,
combining their $\gamma\gamma$ bound with the BES $J/\psi$ radiative decay 
rate~\cite{xibes}, that
for the $\xi(2230)$ stickiness is $>100$, whilst here L3 at LEP~\cite{L3} deduce $>33$.
This would clearly suggest that the $\xi(2230)$, if confirmed, is rich in glue.
But one should be aware that even a pure $q{\overline q}$ state, like the $f_2'(1525)$
has ${\cal S} = 15 \pm 4$. The large value coming from the fact that
its two photon decay rate is smaller than that of the $f_2(1270)$, just because it contains
{\it strange} as opposed to {\it up} quarks.
Because the scalar glueball candidates mix with quark states,
either directly or through their common hadronic decay modes, they
will inevitably have two photon couplings comparable to those of $q{\overline q}$
mesons.

\begin{figure}[t]
\begin{center}
~\epsfig{file=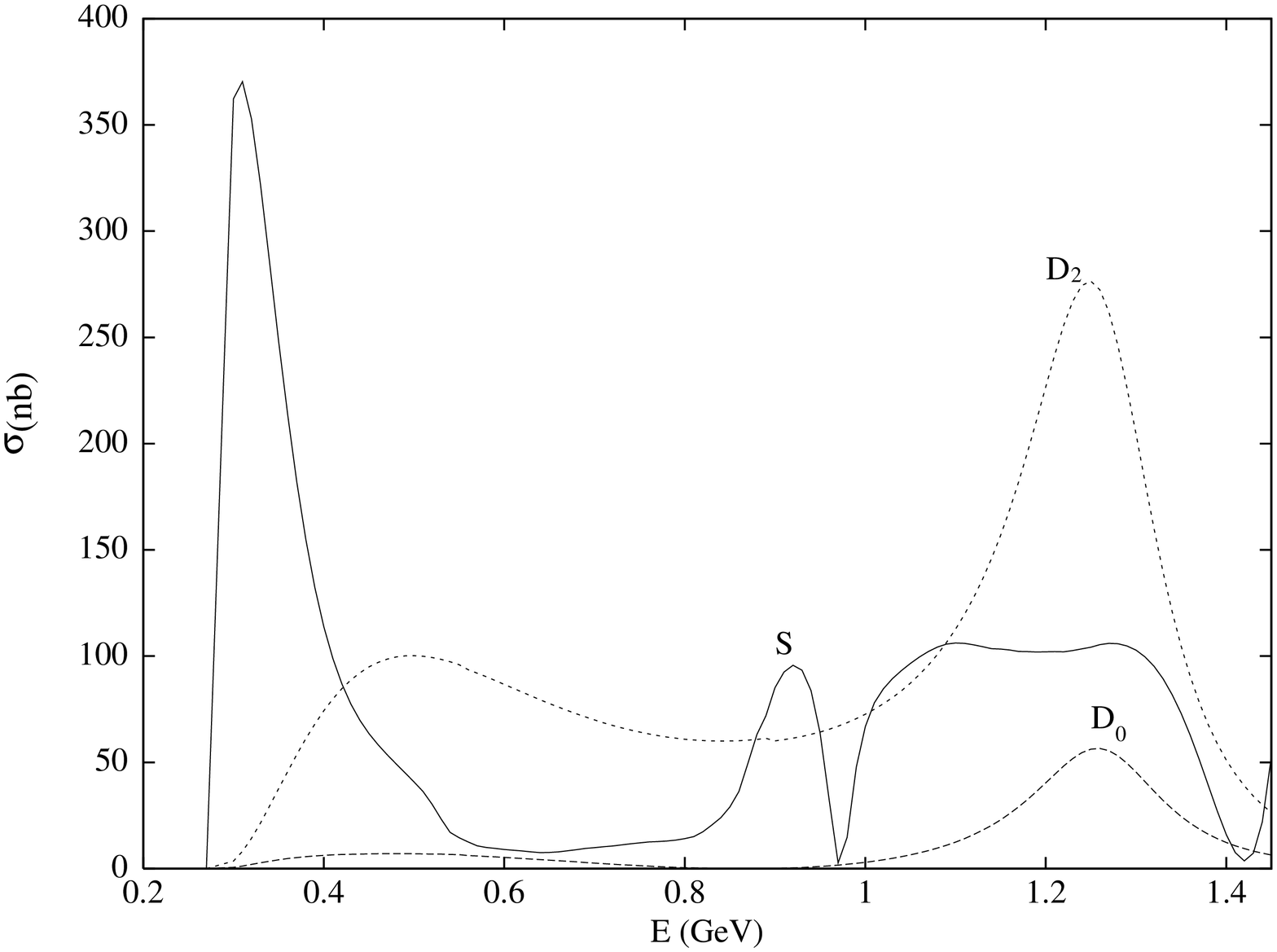,angle=-90,width=13.5cm}
\captive{Integrated I=0 $\gamma \gamma \to \pi \pi$ cross-section in its
spin and helicity components (labelled by $J_{\lambda}$) as a function of
energy (for the {\it dip} solution~\cite{elenaphot}).
 Notice the pronounced peak in the $S$--wave at threshold, produced by the Born amplitude, 
 where it dominates the
$\pi ^{+}\pi ^{-}$ cross-section, and the dip corresponding to the
$f_0(980)$ resonance.} 
\end{center}
\vspace{-5mm}
\end{figure}

To see what we expect for these and for the well known $f_2(1270)$, let us consider the
experimental situation for $\gamma\gamma\to\pi\pi$.
I first want to stress that though we have a clear signal for the $f_2(1270)$
in both the $\pi^+\pi^-$ and $\pi^0\pi^0$ channels~\cite{ggrev}, its two photon width cannot be
reliably extracted without assumptions.
This is because, in $e^+e^-$ colliders, $\gamma\gamma$ data inevitably have limited angular coverage
and decomposing the measured cross-sections into components with definite spins
 and helicities, is not straightforward.
 Only by assuming, for instance, that
the cross-section at 1270 MeV is pure $D$--wave with helicity two 
have results for the width been deduced~\cite{marsiske,boyer}. 
What is really needed is a true Amplitude Analysis. By the use of key theoretical
constraints, that only apply to such a special $\gamma\gamma$ process,
can  a partial wave separation be achieved. Because of the large $I=2$ $S$--wave
component at low energies, produced by the Born amplitude, data on
both the charged and neutral channels are crucial in accomplishing this.
A new Amplitude Analysis has just been completed by Elena Boglione and 
myself~\cite{elenaphot}
covering both the older experiments of Mark II at SLAC~\cite{boyer} and 
Crystal Ball 
at DESY~\cite{marsiske}, to which are added
results from CELLO~\cite{behrend} and from the extended Crystal Ball dataset~\cite{bienlein}.
Present data allow  two distinct  solutions, one of which
I will discuss here as an example.  We find the integrated partial wave cross-sections
 in the isoscalar channel shown in Fig.~9  (for what we call the {\it dip} solution).
While helicity 2 dominates the $D$--wave signal, the
$D_0$ component is far from negligible. Adding these gives
$\Gamma(f_2\to\gamma\gamma) = (2.64 \pm 0.34)$ keV, whether evaluated from the
residue of the pole, or more simply from the \lq\lq peak" height.
For the $f_0(980)$, which appears in this solution as a dip structure, just as in
$\pi\pi\to\pi\pi$, and for the broad $f_0(400-1200)$
we find
\setcounter{equation}{3}
\begin{eqnarray}
\nonumber
\Gamma(f_0(980)\to\gamma\gamma)&=&(0.32 \pm 0.05)\,{\rm keV}\quad ,\\
\Gamma(f_0(400-1200)\to\gamma\gamma)&=&(~4.7 \pm 1.5~)\;\; {\rm keV}\quad .
\end{eqnarray}
For the \lq\lq peak" solution, see Ref.~\cite{elenaphot} for the analogous numbers.
That solution gives $\Gamma(f_2\to\gamma\gamma) = (3.04 \pm 0.38)$ keV.
In the discussion below, where exact results are not required,
I will take $\Gamma(f_2\to\gamma\gamma) \simeq 2.8$ keV as a guide.

Now if the scalars formed a nonet that mirrored the tensors, Li {\it et al.}~\cite{lirel}
have shown the simple 15/4 relation of the non-relativistic quark model has
sizeable corrections giving the predictions of Table 1.
To this have been added other estimates, for instance for a $K{\overline K}$ molecular 
state~\cite{barnesgg}.
Notice the dependence of the predictions on the model of the make-up
of the states. Though the $\gamma\gamma$ width
does depend on the charges of the constituents as in Eq.~(3), it crucially depends
on the probability that these constituents annihilate --- in potential model
language, it depends on the wave-function at the origin. So though the
fourth power of the charge of the constituents is far greater for
a $K{\overline K}$ molecule than for a simple $s{\overline s}$ bound state, the molecule is a much more diffuse
system, so that the probability of annihilation to form photons is 
strongly suppressed~\cite{barnesgg}.
\begin{table}[t]
\begin{center}
\begin{tabular}{|l||c|c|c|c|}
\cline{2-5}
\multicolumn{1}{c|}{~~}
&\rule[-0.6cm]{0cm}{8mm}  $n \overline n$
& $s \overline s$ & 
$K \overline K$ & bare $gg$
\rule[0.4cm]{0cm}{8mm}  \\ 
\hline \hline 
\rule[-0.8cm]{0cm}{8mm}   
$\Gamma(0^{++}\to\gamma\gamma)$ & ~~~4.5~~~ & ~~~0.4~~~ & ~~~0.6~~~ & ~~~0~~~~ 
\rule[0.4cm]{0cm}{8mm}   \\
\hline 
\end{tabular}
\captive{Two-photon partial widths in $keV$ predicted for a
conventional $q \overline q$ nonet of 
scalar states in the 1--1.3 GeV region, for a $K \overline K$ molecule and a bare glueball.}
\end{center}
\vspace{-5mm}
\end{table}

A bare glueball would not couple to photons (Table~1). Turning on quarks,
{\it unquenching} in lattice-speak, brings a coupling to ordinary hadrons
and the chance to appear in two photon processes.  Exactly what the width should be
depends critically on the mixing scheme and our ability to calculate from this.
If a state is a mixture of a bound state of quarks and of hadronic dressing,
like the $a_0(980)$ of Eq.~(2), how to proceed with such a calculation is not immediately obvious.
This is a target for the near future.

 A similar dilemma arises for the mixed glue states.
Close {\it et al.}~\cite{closef} determine the ratio of the $\gamma\gamma$ width of the
$f_0(1370)$, $f_0(1500)$ and $f_0(1710)$. Let us normalise this so that
$\Gamma(f_0(1370)\to \gamma\gamma)$ is 4 keV (cf Eq.~(4) and Table 1).
Then in their mixing scheme, described in Sect.~3, 
where the $f_0(1500)$ is largely gluish,
they predict its $\gamma\gamma$ width to be $\sim 0.3$ keV. While in the
Lee--Weingarten scheme~\cite{leewein}, where the $f_0(1500)$ is predominantly $s\overline s$,
the width is only $\sim 0.06$ keV.  In both cases, they expect
$\Gamma(f_0(1710)) \simeq 1$ keV. 

The model-dependence of such predictions is dramatically illustrated by the
recent calculations of Jaminon and van den Bosche~\cite{jaminon}
 within a similar mixing scenario and of Burakovsky and Page~\cite{page}. 
 Jaminon {\it et al.} find (with their vacuum
gluon condensate parameter $\chi_0 = 250$ MeV) that
$\Gamma(f_0(1370)\to\gamma\gamma) \simeq 3.1$ keV,
$\Gamma(f_0(1500)\to\gamma\gamma) \simeq 0.13$ keV,
$\Gamma(f_0(1710)\to\gamma\gamma) \simeq 0.018$ keV, quite different 
in magnitude from Close {\it et al.}~\cite{closef}.
Burakovsky and Page~\cite{page} also find  large differences between 
mixing schemes with $\Gamma(f_0(1500)\to\gamma\gamma) \simeq 0.7$ keV,
$\Gamma(f_0(1710)\to\gamma\gamma) \simeq 7$ keV in one case and 0.1 keV and 0.01 keV,
respectively, in their second case, with
 $\Gamma(f_0(1370)\to \gamma\gamma)$  again 4 keV.  
From QCD sum-rules, Narison~\cite{narison} predicts that even
an unmixed scalar glueball at 1500 MeV would have a radiative width of
 0.2--1.8 keV, the larger value being not very {\it sticky}!

There are thus two challenges : one theoretical, the other experimental.
Since the underlying scalars, whether made of quarks or glue --- the bare states
--- are so different from dressed hadrons, we must be able
to compute these hadronic properties reliably, if we are going to make sense of 
experiment. ALEPH have presented an upper bound  of 0.17 keV
on the two photon width of the $f_0(1500)$~\cite{ALEPH}.  To go further one must be able
to study the $2\pi$, $4\pi$ and $K\overline K$ final states in detail
simultaneously, if we are to extract a true scalar signal from under the
dominant spin 2 effects in this region : a demanding task, but one that
is essential, if we are to uncover the dressed hadrons and so find the naked
states beneath. Only then will we be certain about which are glueballs.

\section*{Acknowledgements}
It is a pleasure to thank G\"oran Jarlskog and Torbj\"orn Sj\"ostrand for having organised
such a stimulating meeting in splendid surroundings : a meeting that
illustrated the richness of the world to be seen by photons.
I acknowledge travel support from the EEC-TMR Programme, Contract No.
CT98-0169, EuroDA$\Phi$NE.

\end{document}